\documentclass[aps,reprint]{revtex4}
\usepackage[dvips]{graphicx}
\usepackage{amsmath}
\usepackage{multirow}
\begin{document}

\title{Phase space factors for double-$\beta$ decay} 

\author{J. Kotila}
\email{jenni.kotila@yale.edu}
\affiliation{Center for Theoretical Physics, Sloane Physics Laboratory, Yale University, New Haven, Connecticut, 06520-8120, USA}
\author{F. Iachello}
\email{francesco.iachello@yale.edu}
\affiliation{Center for Theoretical Physics, Sloane Physics Laboratory, Yale University, New Haven, Connecticut, 06520-8120, USA}

\begin{abstract}
A complete and improved calculation of phase space factors (PSF) for $2\nu\beta\beta$ and $0\nu\beta\beta$ decay is presented. The calculation makes use of exact Dirac wave functions with finite nuclear size and electron screening and includes life-times, single and summed electron spectra, and angular electron correlations.
\end{abstract}

\maketitle

\section{Introduction}
Double-$\beta$ decay is a process in which a nucleus $(A,Z)$ decays to a nucleus $(A,Z\pm2)$ by emitting two electrons (or positrons) and, usually, other light particles
\begin{equation}
(A,Z)\rightarrow(A,Z\pm 2) + 2e^{\mp} + \text{anything}.
\end{equation}
Double-$\beta$ decay can be classified in various modes according to the various types of particles emitted in the decay.
\begin{figure}[h]
\includegraphics[width=0.95\linewidth]{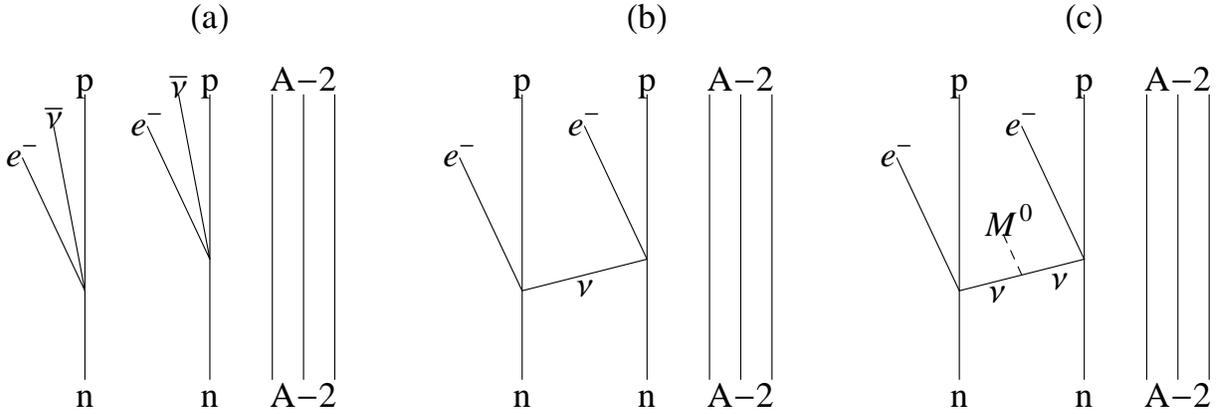} 
\caption{\label{feynman}Double-$\beta$ decay mechanism for (a) two-neutrino, (b) neutrinoless and (c) neutrinoless decay with Majoron emission.}
\end{figure}
For $\beta^-\beta^-$, the process $2\nu\beta\beta$, Fig.~\ref{feynman}a,
\begin{equation}
(A,Z)\rightarrow(A,Z+2) + 2e^{-} + 2\bar{\nu}
\end{equation}
is allowed by the standard model and expected to occur with calculable probability.
In recent years, the process $0\nu\beta\beta$, Fig.~\ref{feynman}b, 
\begin{equation}
(A,Z)\rightarrow(A,Z+2) + 2e^{-}
\end{equation}
has become of great interest, due to the discovery of neutrino oscillation \cite{SUP01, SNO2002, KAM2003}. The process is of utmost importance for obtaining the neutrino mass since its decay probability is proportional to the square of the average neutrino mass $\langle m_{\nu} \rangle$. A third process has been also considered, $0\nu\beta\beta M$, Fig.~\ref{feynman}c,
\begin{equation}
(A,Z)\rightarrow(A,Z+2) + 2e^{-}+M^0
\end{equation}
in which a massless Nambu-Goldstone boson, called a Majoron, is emitted. However, most of the interest in this mode has disappeared in recent years and hence it will not be considered here. For $\beta^+\beta^+$ decay, the corresponding modes $2\nu\beta\beta$, $0\nu\beta\beta$, are
\begin{equation}
\begin{split}
(A,Z)&\rightarrow(A,Z-2) + 2e^{+} +2\nu \\
(A,Z)&\rightarrow(A,Z-2) + 2e^{+}.
\end{split}
\end{equation}
In this case, there are also the competing modes in which either one or two electrons are captured from the electron cloud, $2\nu\beta EC$, $2\nu ECEC$, $0\nu\beta EC$, $0\nu ECEC$.

For processes allowed by the standard model ($2\nu\beta\beta$, $2\nu\beta EC$, $2\nu ECEC$) the half-life can be, to a good approximation, factorized in the form
\begin{equation}
\left[\tau^{2\nu}_{1/2}\right]^{-1}=G_{2\nu}|M_{2\nu}|^2,
\end{equation}
where $G_{2\nu}$ is a phase space factor and $M_{2\nu}$ the nuclear matrix element. For processes not allowed by the standard model the half-life can be factorized as 
\begin{equation}
\left[\tau^{0\nu}_{1/2}\right]^{-1}=G_{0\nu}|M_{0\nu}|^2 \left| f(m_i,U_{ei})\right|^2,
\end{equation}
where $G_{0\nu}$ is a phase space factor, $M_{0\nu}$ the nuclear matrix element and $f(m_i, U_{ei})$ contains physics beyond the standard model through the masses $m_i$ and mixing matrix elements $U_{ei}$ of neutrino species. For both processes, two crucial ingredients are the phase space factors and the nuclear matrix elements. Recently, we have initiated a program for the evaluation of both quantities. For the nuclear matrix elements we have developed an approach based on the microscopic interacting boson model (IBM-2) and presented some results in \cite{barea}. Additional preliminary results have been presented in \cite{barea11,iac11} and will be discussed in a forthcoming publication \cite{barea12}. In this article, we concentrate on phase space factors.

A general theory of phase space factors in DBD was developed years ago by Doi \textit{et al.} \cite{doi81,doi83a} following previous work of Primakoff and Rosen \cite{primakoff} and Konopinski \cite{konopinski}. It was reformulated by Tomoda \cite{tom91} whose work we follow here. Tomoda also presented results in a selected number of nuclei. These results were obtained by approximating the electron wave functions at the nuclear radius and without inclusion of electron screening. In this article we take advantage of some recent developments in the numerical evaluation of Dirac wave functions and in the solution of the Thomas-Fermi equation to calculate more accurate phase space factors for double-$\beta$ decay in all nuclei of interest. Our results are of particular interest in heavy nuclei, $\alpha Z$ large, where relativistic and screening corrections play a major role. Studies similar to ours were done for single-$\beta$ decay in the 1970's \cite{single}.
In this article we report results for $\beta^-\beta^-$, which at the moment is the most promising decay mode. In a subsequent publication, we will present results for $\beta^+\beta^+$, $\beta^+EC$, $ECEC$, which is very recently attracting some attention \cite{positron}.

\section{Electron wave functions}
The key ingredients for the evaluation of phase space factors in single- and double-$\beta$ decay are the (scattering) electron wave functions. (For EC the bound wave functions.) The general theory of relativistic electrons can be found e.g., in  the book of Rose \cite{rose}. We use, for $\beta$ decay,  positive energy Dirac central field wave functions,
\begin{equation}
\psi_{\epsilon\kappa\mu}(\mathbf{r})=\left(
\begin{array}{c}
g_{\kappa}(\epsilon,r)\chi_{\kappa}^{\mu}\\
if_{\kappa}(\epsilon,r)\chi_{-\kappa}^{\mu},
\end{array}
\right),
\end{equation}
where $\chi_{\kappa}^{\mu}$ are spherical spinors and $g_{\kappa}(\epsilon,r)$ and $f_{\kappa}(\epsilon,r)$ are radial functions, with energy $\epsilon$, depending on the relativistic quantum number $\kappa$ defined by $\kappa=(l-j)(2j+1)$. Given an atomic potential $V(r)$ the functions $g_{\kappa}(\epsilon,r)$ and $f_{\kappa}(\epsilon,r)$ satisfy the radial Dirac 
equations:
\begin{equation}
\begin{split}
\frac{dg_{\kappa}(\epsilon,r)}{dr}&=\frac{\kappa}{r}g_{\kappa}(\epsilon,r)+\frac{\epsilon-V+m_ec^2}{c\hbar}f_{\kappa}(\epsilon,r), \\
\frac{df_{\kappa}(\epsilon,r)}{dr}&=-\frac{\epsilon-V-m_ec^2}{c\hbar}g_{\kappa}(\epsilon,r)+\frac{\kappa}{r}f_{\kappa}(\epsilon,r). 
\end{split}
\end{equation}
The electron scattering wave function, denoted here by $e_{s}(\epsilon,r)$, where $s$ is the projection of the spin, can then be expanded in terms of spherical waves as
\begin{equation}
e_{s}(\epsilon,\mathbf{r})=e^{S_{1/2}}_{s}(\epsilon,\mathbf{r})+e^{P_{1/2}}_{s}(\epsilon,\mathbf{r})+e^{P_{3/2}}_{s}(\epsilon,\mathbf{r})+...
\end{equation}
where
\begin{equation}
\begin{split}
e^{S_{1/2}}_{s}(\epsilon ,\mathbf{r})&=\left(\begin{array}{c}
g_{-1}(\epsilon ,r)\chi_s\\
f_{1}(\epsilon,r)(\hat{\mathbf{p}}\cdot \vec{\sigma})\chi_s
\end{array}
\right)\\
e^{P_{1/2}}_{s}(\epsilon ,\mathbf{r})&=\left(\begin{array}{c}
ig_{1}(\epsilon,r)(\hat{\mathbf{r}}\cdot \vec{\sigma})(\hat{\mathbf{p}}\cdot \vec{\sigma})\chi_s\\
-if_{-1}(\epsilon,r)(\hat{\mathbf{r}}\cdot \vec{\sigma})\chi_s
\end{array}
\right)\\
e^{P_{3/2}}_{s}(\epsilon,\mathbf{r})&=\left(\begin{array}{c}
ig_{-2}(\epsilon,r)[3(\hat{\mathbf{r}}\cdot \hat{\mathbf{p}})-(\hat{\mathbf{r}}\cdot \vec{\sigma})(\hat{\mathbf{p}}\cdot \vec{\sigma})]\chi_s\\
if_{2}(\epsilon,r)[3(\hat{\mathbf{r}}\cdot \hat{\mathbf{p}})(\hat{\mathbf{p}}\cdot \vec{\sigma})-(\hat{\mathbf{r}}\cdot \vec{\sigma})]\chi_s
\end{array}
\right).
\end{split}
\end{equation}
The large and small components $g_{\kappa}(\epsilon,r)$ and $f_{\kappa}(\epsilon,r)$, respectively, with $\epsilon=\sqrt{(m_ec^2)^2+(pc)^2}$ of the radial wave functions are normalized so that they asymptotically oscillate with
\begin{equation}
\left(
\begin{array}{c}
g_{\kappa}(\epsilon,r)\\
f_{\kappa}(\epsilon,r)
\end{array}
\right)\sim
e^{-i\delta_{\kappa}}\frac{\hbar}{pr} 
\left(
\begin{array}{c}
\sqrt{\frac{\epsilon+m_ec^2}{2\epsilon}}\sin(kr-l\frac{\pi}{2}-\eta\ln(2kr)+\delta_{\kappa} ) \\
\sqrt{\frac{\epsilon-m_ec^2}{2\epsilon}}\cos(kr-l\frac{\pi}{2}-\eta\ln(2kr)+\delta_{\kappa} )
\end{array}
\right),
\end{equation}
where
\begin{equation}
k\equiv\frac{p}{\hbar}=\frac{\sqrt{\epsilon^2+(m_ec^2)^2}}{c\hbar}
\end{equation}
is the electron wave number, $\eta=Ze^2/\hbar v$ is the Sommerfeld parameter and $\delta_k$ is the phase shift. (For the neutrino wave functions appearing in the $2\nu$ decay mode the limit $Z\rightarrow 0$ is taken, in which case the wave functions become the spherical Bessel functions.)

The radial wave functions are evaluated by means of the
subroutine package RADIAL \cite{sal95}, which implements a robust solution method
that avoids the accumulation of truncation errors. This is done by solving the radial equations by using a piecewise exact power series expansion of the radial functions, which then are summed up to the prescribed accuracy so that truncation errors can be completely avoided. The input in the package is the potential $V$.  This potential is primarily the Coulomb potential of the daughter nucleus with charge $Z_d$, $V(r)=-Z_d(\alpha\hbar c)/r$. As in the case of single-$\beta$ decay \cite{single} we include nuclear size corrections and screening.

The nuclear size corrections are taken into account by an uniform charge distribution  in a sphere of radius $R=r_0A^{1/3}$ with $r_0=1.2$ fm, i.e. 
\begin{equation}
V(r)=\left[\begin{array}{lll}
-\frac{Z_{d}(\alpha\hbar c)}{r}&, &r\geq R\\
-Z_{d}(\alpha\hbar c)\left(\frac{3-(r/R)^2}{2R}\right)&, &r< R
\end{array}\right].
\end{equation}
The introduction of finite nuclear size has also the advantage that the singularity at the origin in the solution of the Dirac equation is removed. (Other charge distributions, for example a Woods-Saxon distribution, can be used if needed.)

The contribution of screening to the phase space factors was extensively investigated in single-$\beta$ decay \cite{wil70, buh65}. The screening potential is of order $V_S\propto Z_d^{4/3}\alpha^2$ and thus gives a contribution of order $\alpha=1/137$ relative to the pure Coulomb potential $V_C\propto Z_d \alpha$. We take it into account by using the Thomas-Fermi approximation.  The Thomas-Fermi function $\varphi(x)$, solution of the Thomas-Fermi equation
\begin{equation}
\label{tf}
\frac{d^2\varphi}{dx^2}=\frac{\varphi^{3/2}}{\sqrt{x}}
\end{equation}
with $x=r/b$  and
\begin{equation}
b=\frac{1}{2}\left(\frac{3\pi}{4}\right)^{2/3}\frac{\hbar ^2}{m_ee^2}Z_d^{-1/3}\simeq 0.8853a_0Z_d^{-1/3},
\end{equation}
where $a_0$ is the Bohr radius, is obtained by solving Eq.~(\ref{tf}) for a point charge $Z_d$ with boundary conditions
\begin{equation}
\begin{split}
\label{bound1}
\varphi(0)&=1,\\
\varphi(\infty)&=\frac{2}{Z_d}.
\end{split}
\end{equation} 
This takes into account the fact that the final atom is a positive ion with charge $+2$. With the introduction of this function, the potential $V(r)$ including screening becomes
\begin{equation}
V(r)\equiv \varphi(r) \times
\left[\begin{array}{lll}
-\frac{Z_d(\alpha\hbar c)}{r} &, &r\geq R\\
-Z_d(\alpha\hbar c)\left(\frac{3-(r/R)^2}{2R}\right) &, &r<R
\end{array}\right].
\end{equation}
This can be rewritten in terms of an effective charge $Z_{\rm{eff}}=Z_d\varphi(r)$ where $Z_{\rm{eff}}$ now depends on $r$. In order to solve Eq.~(\ref{tf}), we use the Majorana method described in \cite{esp02} which is valid both for a neutral atom and a positive ion. The method requires only one quadrature and is thus amenable to a simple solution. It is particularly useful here, since we want to evaluate screening corrections in several nuclei.
The Thomas-Fermi electron density is approximate, especially at  the origin. However, the screening correction is only of order $\alpha$ relative to the Coulomb potential and the error on this small correction is therefore negligible. (A better method would be to do an atomic Hartree-Fock calculation and then fit the result to the expansion 
\begin{equation}
V(r)=(- Z_d(\alpha\hbar c)/r)\sum_i a_i exp(-b_i x), 
\end{equation}
where $x=r/b$ as in Eq.~(\ref{tf}). However, it has been shown in single-$\beta$ decay that this method gives results comparable to the Thomas-Fermi approximation \cite{buh65}, except in very light nuclei, $Z\leq 8$, which we do not discuss here.) We also do not consider radiative corrections to the phase space factors which are of order $\alpha^3$ and thus negligible to the order of approximation we consider in this article.

\begin{figure}[h]
\includegraphics[width=0.75\linewidth]{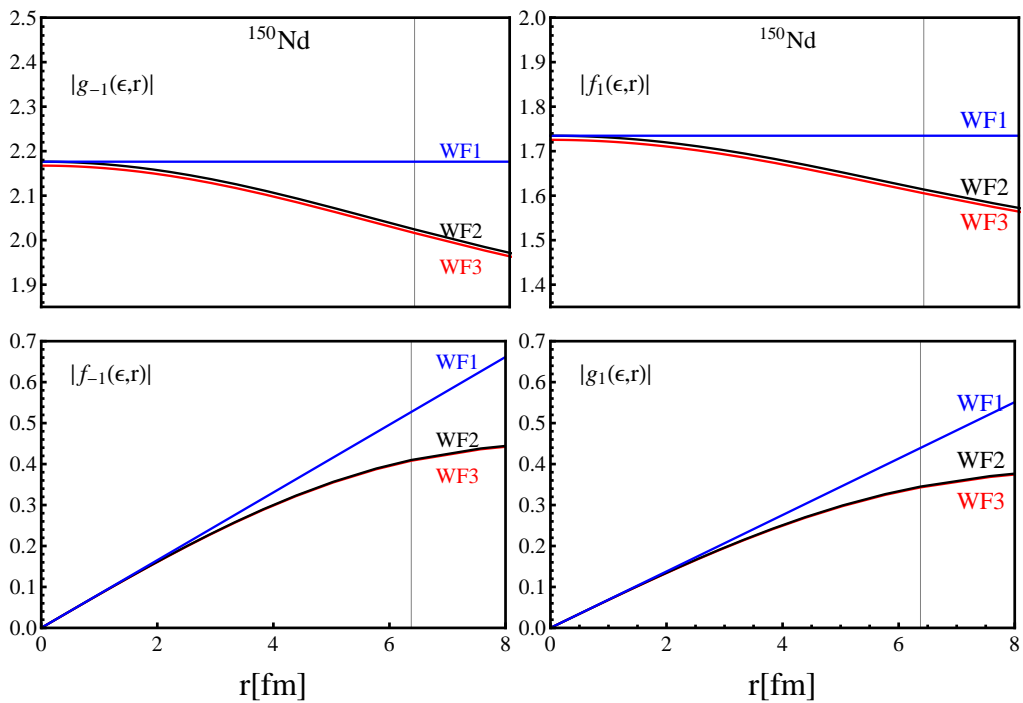} 
\caption{\label{wf}Electron radial wave functions $g_{-1}(\epsilon,r)$, $f_{-1}(\epsilon,r)$ (left panel) and $f_{1}(\epsilon,r)$, $g_{1}(\epsilon,r)$ (right panel) for $Z_d=62$, $\epsilon=2.0$ MeV and $R=6.38$ fm (vertical line). The notations WF1, WF2, and WF3 correspond to leading finite size Coulomb, exact finite size Coulomb and  exact finite size Coulomb with electron screening, respectively.}
\end{figure}

In order to show the improvement in our calculation as compared with the approximate solution used in the literature we show in Fig.~\ref{wf} a comparison of the radial wave functions for $^{150}$Nd decay, $Z_d=62$, at $\epsilon=2.0$ MeV. 

\section{Phase space factors in double-$\beta$ decay}
\subsection{Two neutrino double-$\beta$ decay}
\begin{figure}[h]
\includegraphics[width=0.65\linewidth]{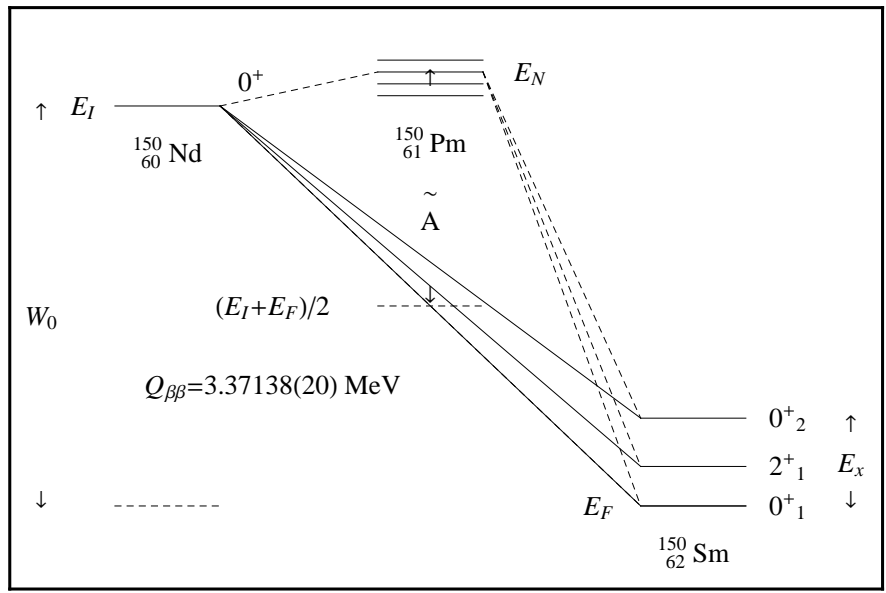} 
\caption{\label{scheme}Notation used in this article. The example is for $^{150}$Nd decay.}
\end{figure}

The $2\nu\beta\beta$ decay, Fig.~\ref{feynman}a, is a second order process in the effective weak interaction. It can be calculated in a way analogous to single-$\beta$ decay. 
Neglecting the neutrino mass, considering only S-wave states and noting that with four leptons in the final state we can have angular momentum $0$, $1$ and, $2$, we see that both $0^+\rightarrow 0^+$ and $0^+\rightarrow 2^+$ decays can occur. We denote by $Q_{\beta\beta}$ the $Q$-value of the decay, by $E_N$ the excitation energy in the intermediate nucleus, and by $\tilde{A}$ the excitation energy with respect to the average of the initial and final ground states, 
\begin{equation}
\tilde{A}=\frac{1}{2}W_0+ E_N-E_I=\frac{1}{2}(Q_{\beta\beta}+2m_ec^2)+ E_N-E_I.
\end{equation}
The situation is illustrated in Fig.~\ref{scheme}.

\subsubsection{$0^+\rightarrow 0^+_1$ $2\nu\beta\beta$-decay}
\label{00 2bb-decay}
The differential rate for $0^+\rightarrow 0^+_1$ $2\nu\beta\beta$-decay is given by 
(\cite{primakoff, konopinski, doi81, doi83a, tom91, haxton})
\begin{equation}
\label{dw2nu}
dW_{2\nu}=\left(a^{(0)}+a^{(1)}\cos \theta_{12}\right)w_{2\nu}d\omega_1d\epsilon_1d\epsilon_2d(\cos \theta_{12})
\end{equation}
where $\epsilon_1$ and $\epsilon_2$ are the electron energies, $\omega_1$ and $\omega_2$ the neutrino energies, $\theta_{12}$ the angle between the two emitted electrons, and 
\begin{equation}
w_{2\nu}=\frac{g_A^4 (G\cos\theta_C)^4}{64\pi^7\hbar}\omega_1^2\omega_2^2(p_1c)(p_2c)\epsilon_1\epsilon_2.
\end{equation}
The quantities $a^{(0)}$ and $a^{(1)}$ are a sum of the contributions of all the intermediate states and depend on the energy $E_N$ of the intermediate state in the odd-odd nucleus and on the nuclear matrix elements $M_{2\nu}$. 
Introducing the short-hand notation 
\begin{equation}
\label{closure}
\begin{split}
\left< K_N\right>&=\frac{1}{\epsilon_1+\omega_1+\left< E_N\right> -E_I}+\frac{1}{\epsilon_2+\omega_2+\left< E_N\right> -E_I},\\
\left< L_N\right>&=\frac{1}{\epsilon_1+\omega_2+\left< E_N\right> -E_i}+\frac{1}{\epsilon_2+\omega_1+\left< E_N\right> -E_I},
\end{split}
\end{equation}
where $\langle E_N \rangle$ is a suitably chosen excitation energy in the odd-odd nucleus, one can write \cite{tom91}, to a good approximation,
\begin{equation}
\label{closure2}
\begin{split}
a^{(0)}&=\frac{1}{4}f_{11}^{(0)}|M_{2\nu}|^2 \tilde{A}^2 \left[\left( \langle K_N\rangle +\langle L_N\rangle \right)^2+\frac{1}{3}\left( \langle K_N\rangle -\langle L_N\rangle \right)^2\right],\\
a^{(1)}&=\frac{1}{4}f_{11}^{(1)}|M_{2\nu}|^2 \tilde{A}^2 \left[\left( \langle K_N\rangle +\langle L_N\rangle \right)^2-\frac{1}{9}\left( \langle K_N\rangle -\langle L_N\rangle \right)^2\right],
\end{split}
\end{equation}
where $M_{2\nu}$ are the nuclear matrix elements and $f_{11}^{(0)}$ and $f_{11}^{(1)}$ are products of radial wave functions. Since Eq.~(\ref{closure2}) is an approximation to the exact expression, which is, however, of crucial importance for the separation of the decay probability into a phase space factor and a nuclear matrix element we have investigated the dependence of $a^{(0)}$ and $a^{(1)}$ on the energy $\left< E_N \right>$. Since $\left< E_N \right>$ appears both in the denominator of Eq.~(\ref{closure2}) through $\left< K_N \right>$ and $\left< L_N \right>$ and in the numerator through $\tilde{A}^2=\left[ W_0/2+\langle E_N \rangle-E_I\right]^2$, the dependence on $\left< E_N \right>$ cancels almost completely, as already remarked years ago by Tomoda \cite{tom91}, and as it is shown by explicit calculation in the following paragraphs.

The functions $f_{11}^{(0)}$ and $f_{11}^{(1)}$ are defined as
\begin{equation}
\label{combwave}
\begin{split}
f^{(0)}_{11}&=|f^{-1-1}|^2+|f_{11}|^2+|{f^{-1}}_{ 1}|^2+|{f_1}^{ -1}|^2,\\
f^{(1)}_{11}&=-2\text{Re}[f^{-1-1}f_{11}^*+{f^{-1}}_{ 1}{f_1}^{ -1*}].
\end{split}
\end{equation}
with
\begin{equation}
\begin{split}
f^{-1-1}&=g_{-1}(\epsilon_1)g_{-1}(\epsilon_2),\\
f_{11}&=f_1(\epsilon_1)f_1(\epsilon_2),\\
{f^{-1}}_{ 1}&=g_{-1}(\epsilon_1)f_1(\epsilon_2),\\
{f_{1}}^{ -1}&=f_1(\epsilon_1)g_{-1}(\epsilon_2).
\end{split}
\end{equation}
The functions $g_{-1}(\epsilon)$ and $f_{1}(\epsilon)$ are obtained from the electron wave functions. We have used several ways to obtain $g_{-1}(\epsilon)$ and $f_{1}(\epsilon)$ following an approach similar to that used in single-$\beta$ decay. We write
\begin{equation}
\begin{split}
g_{-1}(\epsilon)&=\int^\infty_0 w(r)g_{-1}(\epsilon,r)r^2 dr,\\
f_{1}(\epsilon)&=\int^\infty_0 w(r)f_{1}(\epsilon,r)r^2 dr.
\end{split}
\end{equation}
In approximation (I) we use the weighing function $w(r)=\delta(r-R)/r^2$ in which case
\begin{equation}
\label{I}
\begin{split}
g_{-1}(\epsilon)&=g_{-1}(\epsilon,R)\\
f_{1}(\epsilon)&=f_{1}(\epsilon,R)
\end{split}
,\qquad \qquad \qquad \rm{(I)}
\end{equation}
that is the electron wave functions are evaluated at the nuclear radius $r=R$. This is the simplest approximation and is commonly used in single-$\beta$ decay. We adopt it in this article. In approximation (II) we use the weighing function $w(r)=3/R^3$ for $r\leq R$ and $w(r)=0$ for $r>R$ (an uniform distribution of radius $R$). This is not a good approximation, since the inner states cannot decay due to Pauli blocking and the decay occurs at the surface of the nucleus. Nevertheless, it is sometimes used. It essentially amounts to an evaluation of $g_{-1}(\epsilon)$ and $f_{1}(\epsilon)$ at a radius $r=\sqrt{3}R/\sqrt{5}$, as one can show by explicitly evaluating
\begin{equation}
\label{II}
\begin{split}
g_{-1}(\epsilon)&=\frac{3}{R^3}\int^R_0 g_{-1}(\epsilon,r)r^2 dr\\
f_{1}(\epsilon)&=\frac{3}{R^3}\int^R_0 f_{1}(\epsilon,r)r^2 dr
\end{split}
.\qquad \qquad \qquad \rm{(II)}
\end{equation}
The third and most accurate approximation (III) is that in which the weighing function is the square of the wave function, $R_{nl}(r)$, of the nucleon undergoing the decay, 
\begin{equation}
\label{III}
\begin{split}
g_{-1}(\epsilon)&=\int^\infty_0 \left|R_{nl}(r) \right|^2 g_{-1}(\epsilon,r)r^2 dr\\
f_{1}(\epsilon)&=\int^\infty_0 \left|R_{nl}(r) \right|^2f_{1}(\epsilon,r)r^2 dr
\end{split}
.\qquad \qquad \qquad \rm{(III)}
\end{equation}
By using harmonic oscillator wave functions and assuming that only one orbital is involved, the integrals in Eq.~(\ref{III}) can be easily evaluated. The approximation (III) essentially amounts to an evaluation of $g_{-1}(\epsilon)$ and $f_{1}(\epsilon)$ at a radius $\sqrt{\left< r^2\right>_{nl}}$. For harmonic oscillator wave functions
\begin{equation}
R_{nl}(r)=\sqrt{\frac{2n!}{b^3\Gamma(n+l+3/2)}}\left( \frac{r}{b}\right)^le^{-r^2/2b^2}L_n^{l+1/2}(r^2/b^2)
\end{equation}
with 
\begin{equation}
b^2=\frac{\hbar}{M\omega}\simeq1.0 A^{1/3}\rm{fm}^2,
\end{equation}
one has 
\begin{equation}
\left< r^2\right>_{nl}=b^2\left( 2n+l+\frac{3}{2}\right).
\end{equation}
This approximation has the disadvantage that it must be done separately for each nucleus. Since in this paper we are seeking greater generality and do not wish to make a commitment to definite nucleonic orbitals, we make use of approximation (I). However, our computer program is written in such way as to allow the possibility of using Eq.~(\ref{III}) instead of Eq.~(\ref{I}). Also in Sect.~\ref{error} we study in a specific case, $^{110}$Pd, where the transition is between $1g_{9/2}$ and $1g_{7/2}$ orbitals, the error we make by using Eq.~(\ref{I}) instead of Eq.~(\ref{III}).

\begin{figure}[h]
\includegraphics[width=0.65\linewidth ]{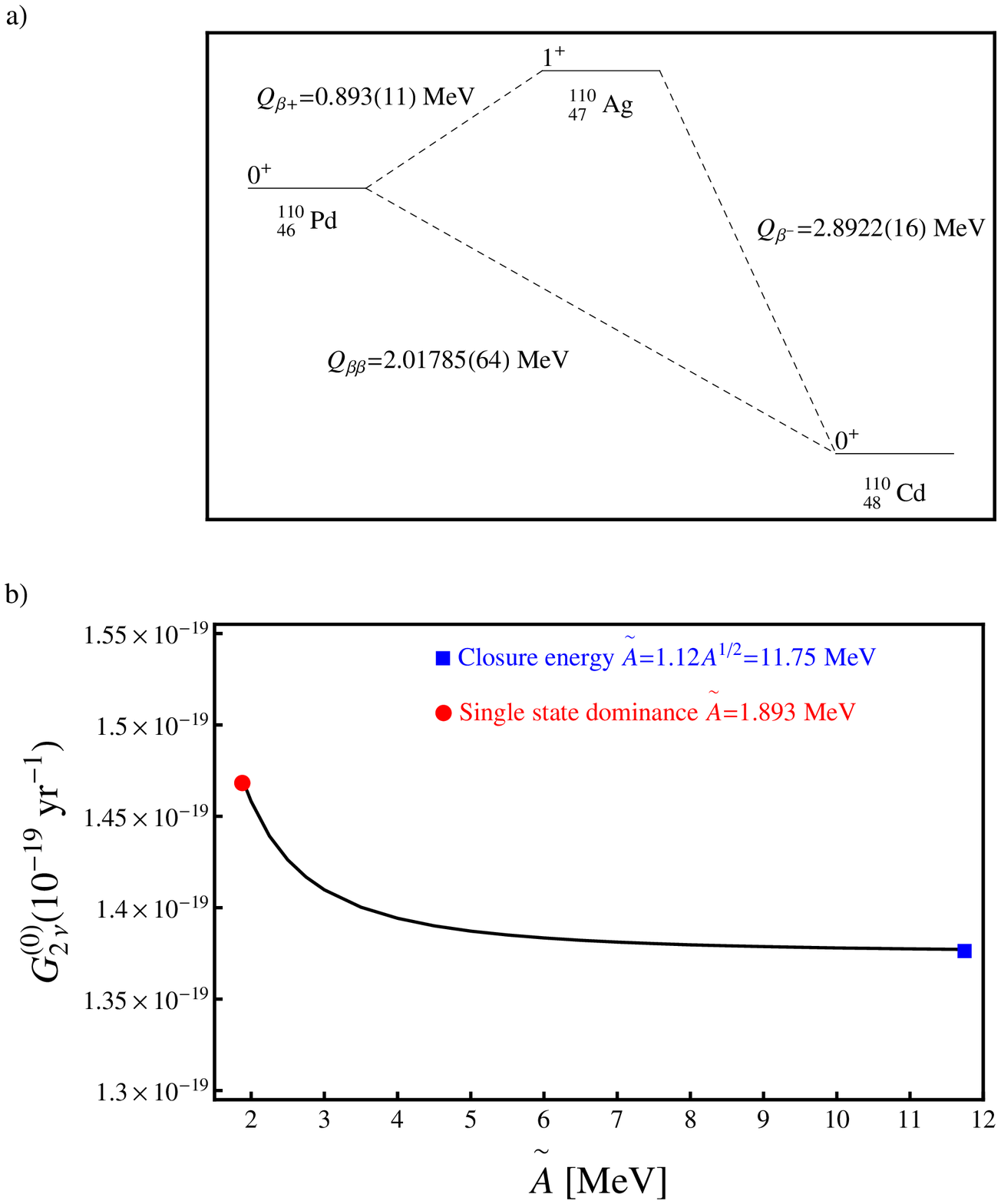} 
\caption{\label{atilde}Panel a) Skeleton of the $^{110}$Pd decay scheme. The ground state of the intermediate $^{110}$Ag nucleus is $1^+$ leading to the lowest possible value for $E_N$ to be $E_{1^+_1}=0.0$~MeV.\\
Panel b) Behaviour of the phase phase factor $G_{2\nu}^{(0)}$ as a function of $\tilde{A}$. The value obtained using single state dominance hypothesis, $\tilde{A}=1.893$ MeV, is denoted by a red circle and the value obtained using $\tilde{A}=1.12\times110^{1/2}$ MeV$=11.75$ MeV is denoted by a blue square.}
\end{figure}

\begin{figure}[h]
\includegraphics[width=0.65\linewidth ]{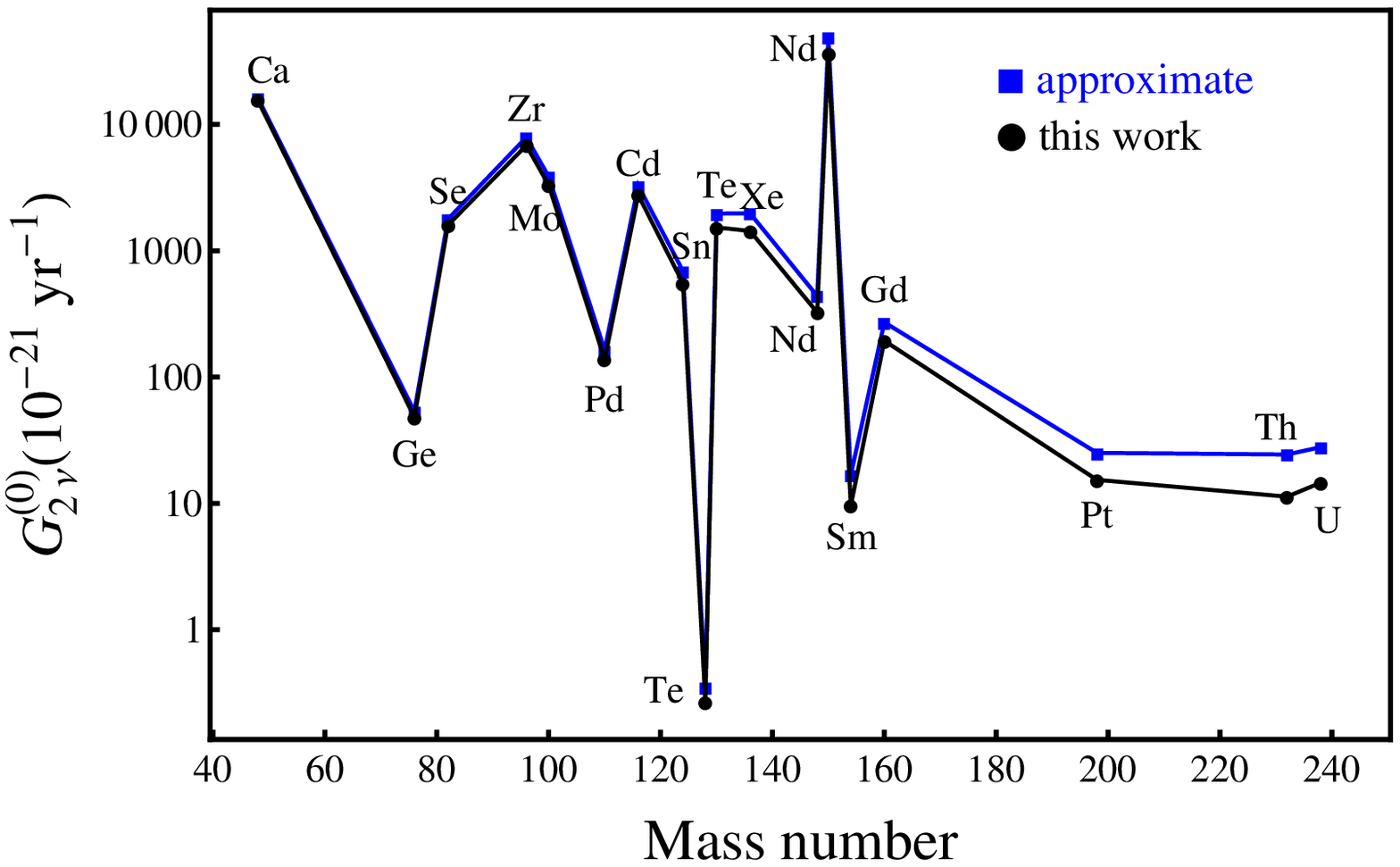} 
\caption{\label{2nuGcomp}Phase space factors $G_{2\nu}^{(0)}$ in units $(10^{-21}$ y$^{-1})$. The label "approximate" refers to the results obtained by the use of approximate electron wave functions. The figure is in semilogarithmic scale.}
\end{figure}

All quantities of interest are obtained by integration of Eq.~(\ref{dw2nu}). In the approximation described above, all quantities are separated into a phase space factor (independent of nuclear matrix elements) and the nuclear matrix elements. The two phase space factors are
\begin{equation}
\begin{split}
F^{(0)}_{2\nu}=\frac{2\tilde{A}^2}{3\ln2}&\int^{Q_{\beta\beta}+m_ec^2}_{m_ec^2}
\int^{Q_{\beta\beta}+m_ec^2-\epsilon_1}_{m_ec^2}
\int^{Q_{\beta\beta}-\epsilon_1-\epsilon_2}_{0}
f^{(0)}_{11}\\
\times &\left( \left< K_N \right> ^2 +\left< L_N \right> ^2 +\left< K_N \right> \left< L_N \right> \right) w_{2\nu}d\omega_1 d\epsilon_2 d\epsilon_1,
\end{split}
\end{equation}

\begin{equation}
\begin{split}
F^{(1)}_{2\nu}=\frac{2\tilde{A}^2}{9\ln2}&\int^{Q_{\beta\beta}+m_ec^2}_{m_ec^2}
\int^{Q_{\beta\beta}+m_ec^2-\epsilon_1}_{m_ec^2}
\int^{Q_{\beta\beta}-\epsilon_1-\epsilon_2}_{0}
f^{(1)}_{11}\\
\times &\left[ 2\left( \left< K_N \right> ^2+\left< L_N \right> ^2\right)
+5\left< K_N \right> \left< L_N \right> \right] w_{2\nu}d\omega_1 d\epsilon_2 d\epsilon_1,
\end{split}
\end{equation}
where $\omega_2$ is determined as $\omega_2=Q_{\beta\beta}-\epsilon_1-\epsilon_2-\omega_1$.
It has become customary to normalize these to the electron mass $m_ec^2$. Also since the axial vector coupling constant $g_A$ is renormalized in nuclei it is convenient to separate it from the phase space factors and define quantities
\begin{equation}
G^{(i)}_{2\nu}=\frac{F^{(i)}_{2\nu}}{g_A^4(m_ec^2)^2}.
\end{equation}
These quantities are then in units of y$^{-1}$. From these, we obtain:\\
(i) The half-life
\begin{equation}
\left[ \tau^{2\nu}_{1/2}\right]^{-1}=G^{(0)}_{2\nu}g_A^4 \left| m_ec^2 M_{2\nu} \right| ^2.
\end{equation}
(ii) The differential decay rate
\begin{equation}
\frac{dW_{2\nu}}{d\epsilon_1}={\cal N}_{2\nu}\frac{dG^{(0)}_{2\nu}}{d\epsilon_1},
\end{equation}
where ${\cal N}_{2\nu}=g_A^4 \left| m_ec^2 M_{2\nu} \right|^2$.\\
(iii) The summed energy spectrum of the two electrons
\begin{equation}
\frac{dW_{2\nu}}{d(\epsilon_1+\epsilon_2-2m_ec^2)}={\cal N}_{2\nu}\frac{dG^{(0)}_{2\nu}}{d(\epsilon_1+\epsilon_2-2m_ec^2)}.
\end{equation}
(iv) The angular correlation between the two electrons
\begin{equation}
\alpha(\epsilon_1)=\frac{dG^{(1)}_{2\nu}/d\epsilon_1}{dG^{(0)}_{2\nu}/d\epsilon_1}.
\end{equation}
We can evaluate the phase space factors $G^{(i)}_{2\nu}$ for any value $\left< E_N \right>$. The dependence of $G^{(0)}_{2\nu}$ on $\tilde{A}=(Q_{\beta\beta}+2m_ec^2)/2+\langle E_N \rangle-E_I$ is shown in Fig.~\ref{atilde}b for the specific case of $^{110}$Pd decay. We see that $G^{(0)}_{2\nu}$ depends mildly on $\tilde{A}$ ($<1\%$) except very close to threshold $\left< E_N \right>=0$, where the dependence is $\sim 7\%$. A similar situation occurs for $G^{(1)}_{2\nu}$. We have done a calculation of $G^{(0)}_{2\nu}$ and $G^{(1)}_{2\nu}$ in the list of nuclei shown in Table \ref{2nuG} with $\tilde{A}$ from Ref.~\cite{haxton} or estimated by the systematics $\tilde{A}=1.12A^{1/2}$ MeV, which approximately represents the energy of the giant Gamow-Teller resonance in the intermediate odd-odd nucleus. The obtained $G^{(0)}_{2\nu}$ values are also shown in Fig.~\ref{2nuGcomp} where they are compared with previous calculations \cite{boe92}. These values of $\tilde{A}$ are those estimated in the closure approximation and should be combined with the closure matrix elements 
\begin{equation}
M_{2\nu}\simeq \left(\frac{g_V}{g_A} \right)^2\frac{M_{2\nu}^{F}}{\tilde{A}^F}-\frac{M_{2\nu}^{GT}}{\tilde{A}^{GT}},
\end{equation}
where $M_{2\nu}^{F}=\langle 0^+_F |\sum_{nn'}\tau_n\tau_{n'}|0^+_I\rangle$ and $M_{2\nu}^{GT}=\langle 0^+_F |\sum_{nn'}\tau_n\tau_{n'}\vec{\sigma}_n\cdot\vec{\sigma}_{n'}|0^+_I\rangle$. Here $\tilde{A}^F$ is the closure energy for $0^+$ states in the odd-odd intermediate nucleus and it can be approximately taken as the energy of the isobaric analogue state.

\begin{figure}[h]
\includegraphics[width=1.0\linewidth ]{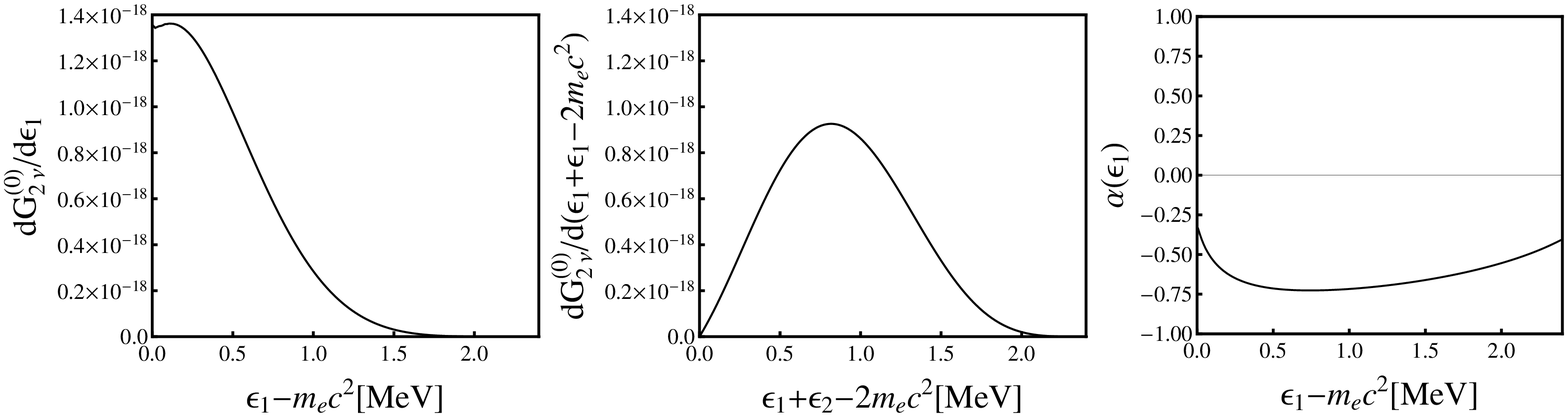} 
\caption{\label{Xe}Single electron spectra (left panel), summed energy spectra (middle panel) and angular correlations between two outgoing electrons (right panel) for the $^{136}$Xe $\rightarrow ^{136}$Ba $2\nu\beta\beta$-decay. The scale in the left and middle panels should be multiplied by ${\cal N}_{2\nu}$ when comparing with experiment.}
\end{figure}

\begin{figure}[h]
\includegraphics[width=1.0\linewidth]{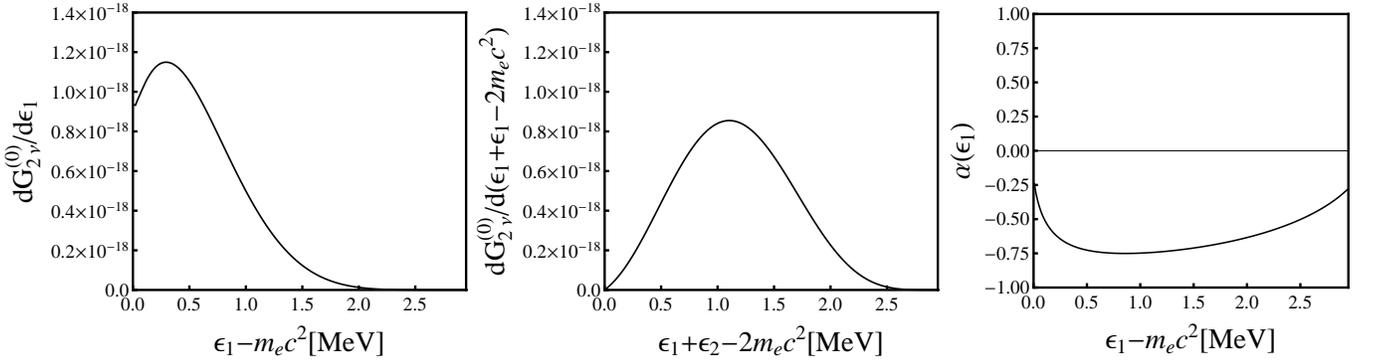} 
\caption{\label{Se}Same as Fig.~\ref{Xe} for the $^{82}$Se $\rightarrow ^{82}$Kr $2\nu\beta\beta$-decay.}
\end{figure}

\begin{figure}[h]
\includegraphics[width=1.0\linewidth]{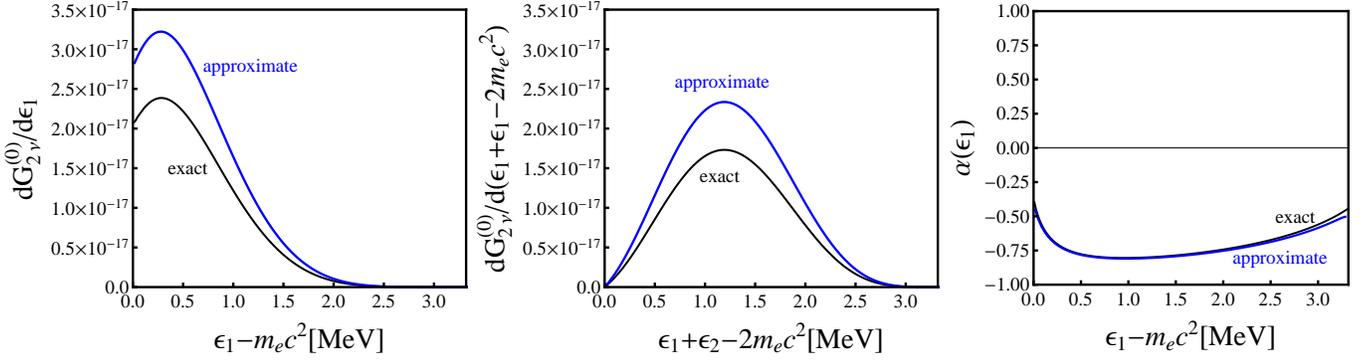} 
\caption{\label{diff}Same as Fig.~\ref{Xe} for the $^{150}$Nd $\rightarrow ^{150}$Sm $2\nu\beta\beta$-decay. The figure also shows the difference between our "exact" calculation and the previously used approximate calculation.}
\end{figure}

\begin{figure}[h]
\includegraphics[width=0.35\linewidth]{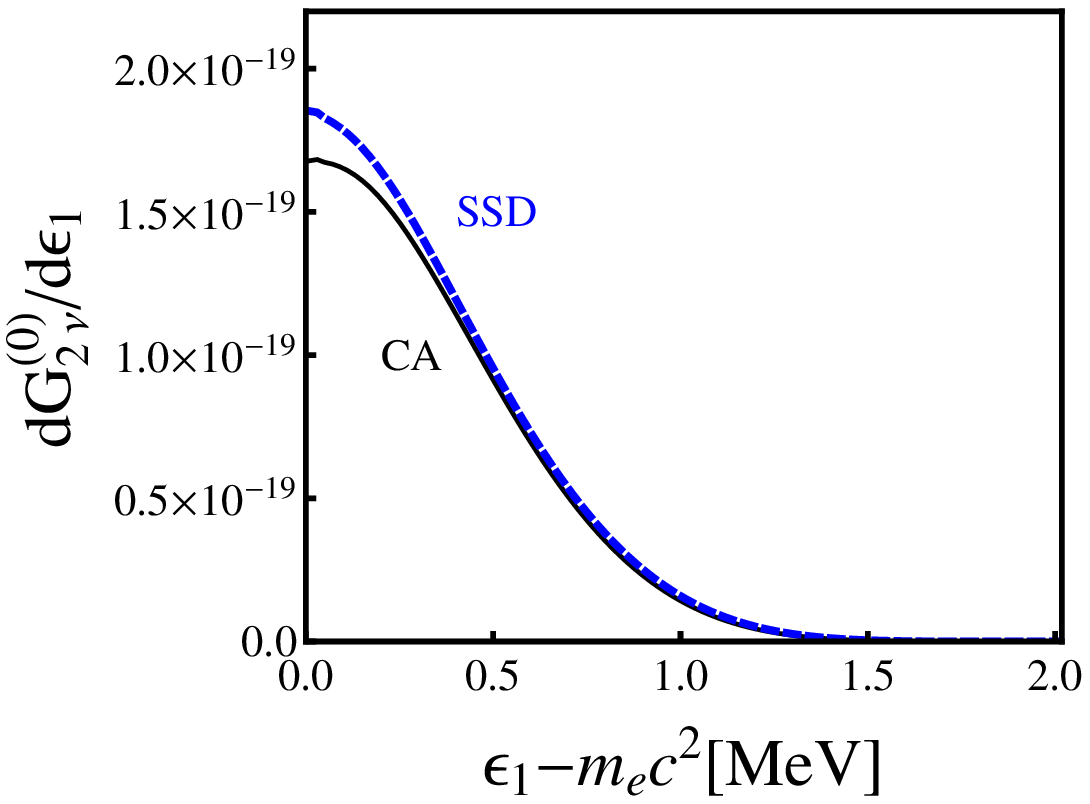} 
\caption{\label{new}Single electron spectra for the $^{110}$Pd $\rightarrow ^{110}$Cd $2\nu\beta\beta$-decay obtained using the two approximations discussed in the text, namely closure approximation and single state dominance hypothesis.}
\end{figure}

In recent years, it has been suggested, that in some nuclei, the lowest $1^+$ intermediate state dominates the decay. This is called the single state dominance hypothesis (SSD) \cite{aba84,gri92,civ98,sim01,dom05}. This situation is likely to occur in $^{96}$Zr, $^{100}$Mo, $^{110}$Pd, and $^{116}$Cd, where protons occupy mostly the $1g_{9/2}$ level and  neutrons mostly the $1g_{7/2}$ level, and in $^{128}$Te, where protons occupy mostly the $2d_{5/2}$ level and  neutrons mostly the $2d_{3/2}$ level, which are spin-orbit partners of each other. In the SSD model the energy $\left< E_N \right>$ is that of the single state $\left< E_N \right>=E_{1^+_1}$. We have done a calculation of $G^{(0)}_{2\nu}$ and $G^{(1)}_{2\nu}$ for the nuclei mentioned above in the SSD case. This is also shown in Table~\ref{2nuG} in columns 3 and 5. In this case, $G^{(0)}_{2\nu}$ and $G^{(1)}_{2\nu}$ should be combined with the matrix elements 
\begin{equation}
M_{2\nu}^{GT}= \frac{\langle 0^+_F ||\tau^+\vec{\sigma}||1^+_1\rangle \langle 1^+_1 ||\tau^+\vec{\sigma}||0^+_I\rangle}{\frac{1}{2}(Q_{\beta\beta}+2m_ec^2)+E_{1^+_1}-E_I}.
\end{equation}
Finally, using our program, one can evaluate the sum
\begin{equation}
\label{full}
\sum_N G^{(i)}_{2\nu, N}\frac{\langle 0^+_F ||\tau^+\vec{\sigma}||1^+_N\rangle \langle 1^+_N ||\tau^+\vec{\sigma}||0^+_I\rangle}{\frac{1}{2}(Q_{\beta\beta}+2m_ec^2)+E_{N}-E_I}
\end{equation}
if the individual GT matrix elements are known from a calculation, and a similar sum for Fermi matrix elements. In this case, there is no separation between $2\nu\beta\beta$ phase space factors and nuclear matrix elements.
\begin{ruledtabular}
\begin{center}
\begin{table}[h]
\begin{tabular}{lccccccc}
Nucleus 	&$G_{2\nu}^{(0)}(10^{-21}$ y$^{-1})$ &${G_{2\nu}^{(0)}}_{\rm{SSD}}(10^{-21}$ y$^{-1})$	&$G_{2\nu}^{(1)}(10^{-21}$ y$^{-1})$ &${G_{2\nu}^{(1)}}_{\rm{SSD}}(10^{-21}$ y$^{-1})$ &$Q_{\beta\beta}$(MeV) &$\tilde{A}$(MeV) &$\tilde{A}_{\rm{SSD}}$(MeV)
\cr \hline 
$^{48}$Ca	&15550. & 		&-11930. &		&4.27226(404) 			&7.717$^h$ 	&\\
$^{76}$Ge	&48.17	&		&-26.97	 &		&2.039061(7)$^a$ 		&9.411$^h$ 	&\\
$^{82}$Se	&1596.  &		&-1075.	 & 		&2.99512(201)	 		&10.08$^h$ 	&\\
$^{96}$Zr	&6816.  &7825.	&-4831.	 &-5477.&3.35037(289)	 		&10.97 		&2.203\\
$^{100}$Mo	&3308.  &4134.	&-2263.	 &-2762.&3.03440(17)$^b$ 		&11.20 		&1.685\\
$^{110}$Pd	&137.7	&146.9	&-79.56  &-84.45.&2.01785(64)$^c$		&11.75 		&1.893\\
$^{116}$Cd	&2764.	&3176.	&-1857.	 &-2108.&2.81350(13)$^d$		&12.06 		&1.875\\
$^{124}$Sn	&553.0	&		&-342.7	 &		&2.28697(153)			&12.47 		&\\
$^{128}$Te	&0.2688 &0.2727	&-0.1047 &-0.1061&0.86587(131)$^e$		&12.53$^h$ 	&1.685\\
$^{130}$Te	&1529.  &		&-993.9  &		&2.52697(23)$^d$		&13.27$^h$ 	&\\
$^{136}$Xe	&1433.  &		&-927.2  & 		&2.45783(37)$^f$		&13.06 		&\\
$^{148}$Nd	&324.8 	&		&-195.5	 &		&1.92875(192)			&13.63 		&\\
$^{150}$Nd	&36430. &		&-26860. &		&3.37138(20)$^g$		&13.72 		&\\
$^{154}$Sm	&9.591	& 		&-4.816	 &		&1.21503(125)			&13.90 		&\\
$^{160}$Gd	&193.8 	&		&-114.2	 &		&1.72969(126)			&14.17 		&\\
$^{198}$Pt	&15.36 	&		&-8.499	 &		&1.04717(311)			&15.76 		&\\
$^{232}$Th	&11.31 	&		&-6.779	 &		&0.84215(246)			&17.06 		&\\
$^{238}$U	&14.57 	&		&-9.543	 &		&1.14498(125)			&17.28 		&\\
\end{tabular}
\caption{\label{2nuG}Phase space factors $G_{2\nu}^{(0)}$ and $G_{2\nu}^{(1)}$ obtained using screened exact finite size Coulomb wave functions. The $Q$-values are taken from experiment ( $^a$ Ref.~\cite{mou10}, $^b$ Ref.~\cite{rah08}, $^c$ Ref.~\cite{fin11}, $^d$ Ref.~\cite{rah11}, $^e$ Ref.~\cite{sci09}, $^f$ Ref.~\cite{red07}, $^g$ Ref.~\cite{kol10}) when available, or from tables of recommended values. $\tilde{A}$ is taken from  $^h$ Ref.~\cite{haxton} or estimated by the systematics, $\tilde{A}=1.12A^{1/2}$ MeV, where A without tilde denotes the mass number. Phase space factors ${G_{2\nu}^{(0)}}_{\rm{SSD}}$ and ${G_{2\nu}^{(1)}}_{\rm{SSD}}$ correspond to values obtained using the SSD model, in which case the used $\tilde{A}_{\rm{SSD}}$ is listed in the last column.}
\end{table}
\end{center}
\end{ruledtabular}
We also have available upon request for all nuclei in Table~\ref{2nuG} the single electron spectra, summed energy spectra and angular correlations between the two outgoing electrons. As examples we show the cases of $^{136}$Xe~$\rightarrow~^{136}$Ba decay, Fig.~\ref{Xe}, of very recent interest to EXO experiment \cite{exo} and the case of  $^{82}$Se~$\rightarrow~^{82}$Kr, Fig.~\ref{Se}, of interest to NEMO experiment \cite{nemo}. The use of our "exact" calculation makes a considerable difference as shown in Fig.~\ref{diff}. For the SSD case there is a difference in the single electron spectra at small energies $\epsilon_1$, as is shown in Fig.~\ref{new} for $^{110}$Pd, and previously emphasized in Refs.~\cite{sim01,dom05}.

\subsubsection{\label{2nu02}$0^+\rightarrow 0^+_2$ $2\nu\beta\beta$-decay}
The decay to the excited $0^+$ state, $0_2^+$ (Fig.~\ref{scheme}), is also of interest. The phase space factor for this decay can be calculated using the formulas of the previous subsection, with $Q_{\beta\beta}$ replaced by
\begin{equation}
Q_{\beta\beta}-E_x(0^+_2)=Q_{\beta\beta}(0^+_2)
\end{equation}
The results of this calculation are shown in Table~\ref{2nuGex}.
\begin{ruledtabular}
\begin{center}
\begin{table}[h]
\begin{tabular}{lcccccc}
Nucleus 	&$G_{2\nu}^{(0)}(10^{-21}$ y$^{-1})$ &${G_{2\nu}^{(0)}}_{\rm{SSD}}(10^{-21}$ y$^{-1})$		&$G_{2\nu}^{(1)}(10^{-21}$ y$^{-1})$ &${G_{2\nu}^{(1)}}_{\rm{SSD}}(10^{-21}$ y$^{-1})$   &$E(0^+_2)$(MeV) &$Q_{\beta\beta}(0^+_2)$(MeV)
\cr \hline 
$^{48}$Ca	&0.3627 	&	&-0.1505 	&	&2.99722(16) 	&1.27504(253)\\
$^{76}$Ge	&0.06978 	&	&-0.02380	&	&1.122283(7) 	&0.916757(167)\\
$^{96}$Zr	&175.4	 	&185.3	&-103.8		&-109.2	&1.14813(7) 	&2.20224(296)\\
$^{100}$Mo	&60.55 		&65.18	&-33.54		&35.89	&1.13032(10) 	&1.90408(27)\\
$^{110}$Pd	&0.004842 	&0.004864	&-0.001371	&-0.001377	&1.47312(12) 	&0.54773(76)\\
$^{116}$Cd	&0.8727	 	&0.8878	&-0.3642	&-0.3701	&1.756864(24) 	&1.056636(154)\\
$^{124}$Sn	&0.01988 	&	&-0.006408	&	&1.657283(22) 	&0.629687(1552)\\
$^{130}$Te	&0.07566 	&	 &-0.02705 	&	&1.79352(11) 	&0.73345(34)\\
$^{136}$Xe	&0.3622	  	&	&-0.1451	&	&1.578990(23) 	&0.878840(393)\\
$^{148}$Nd	&0.009911 	& 	&-0.003339 	&	&1.42446(4) 	&0.50429(196)\\
$^{150}$Nd	&4329.	  	&	&-2934.		&	&0.740382(22) 	&2.630998(222)\\
$^{154}$Sm	&0.01850  	&	&-0.006583	&	&0.6806673(18) 	&0.5343627(12518)\\
$^{160}$Gd	&0.006318 	&	&-0.002178	&	&1.279941(23) 	&0.449749(1283)\\
$^{232}$Th	&0.00004221 &	&-0.00001944&	&0.69142(9)		&0.15073(255)\\
$^{238}$U	&0.0004635  &	&-0.0002289 &	&0.94146(8)		&0.20352(133)\\
\end{tabular}
\caption{\label{2nuGex}Phase space factors $G_{2\nu}^{(0)}$ and $G_{2\nu}^{(1)}$ for decay to the first excited $0^+$ states, $0^+_2$, obtained using screened exact finite size Coulomb wave functions. Phase space factors ${G_{2\nu}^{(0)}}_{\rm{SSD}}$ and ${G_{2\nu}^{(1)}}_{\rm{SSD}}$ correspond to values obtained using the SSD model.}
\end{table}
\end{center}
\end{ruledtabular}

\subsubsection{$0^+\rightarrow 2^+_1$ $2\nu\beta\beta$-decay}
The half-life for $0^+\rightarrow 2^+_1$ $2\nu\beta\beta$-decay is given by equations similar to those of sect.~\ref{00 2bb-decay}
 \cite{molina, doi81, haxton}.
The lepton phase space factor $F^{(0)0^+\rightarrow 2^+_1}_{2\nu}$ is now
\begin{equation}
\begin{split}
F^{(0)0^+\rightarrow 2^+_1}_{2\nu}=\frac{2\tilde{A}^6}{\ln2}&\int^{Q_{\beta\beta}(2^+_1)+m_ec^2}_{m_ec^2}
\int^{Q_{\beta\beta}(2^+_1)+m_ec^2-\epsilon_1}_{m_ec^2}
\int^{Q_{\beta\beta}(2^+_1)-\epsilon_1-\epsilon_2}_{0}
f^{(0)}_{11}\\
\times &\left( \left< K_N \right> -\left< L_N \right> \right) ^2  w_{2\nu}d\omega_1 d\epsilon_2 d\epsilon_1,
\end{split}
\end{equation}
with $Q_{\beta\beta}(2^+_1)=Q_{\beta\beta}-E_x(2^+_1)$, (Fig.~\ref{scheme}), from which the life-time can be calculated
\begin{equation}
\left[ \tau^{2\nu}_{1/2}(0^+\rightarrow 2^+) \right]^{-1}=F^{(0)0^+\rightarrow 2^+}_{2\nu}\left|  M_{2\nu}^{(2^+)} \right| ^2.
\end{equation}
The nuclear matrix elements can be written, in the closure approximation, as
\begin{equation}
M_{2\nu}^{(2^+)}\simeq -\frac{M_{2\nu}^{GT (2^+)}}{\tilde{A}^3}
\end{equation}
where
\begin{equation}
M_{2\nu}^{GT (2^+)}=\langle 2^+_F ||\sum_{nn'}\tau_n\tau_{n'}\left[\vec{\sigma}_n\otimes \vec{\sigma}_{n'}\right]^{(2)}||0^+_I\rangle.
\end{equation}

Since this decay contains the term $\langle K_N \rangle -\langle L_N \rangle$, it is suppressed, due to cancellations, and it will not be considered further. Also, other models (SSD, no-closure) can be used, if needed.

\subsection{Neutrinoless double-$\beta$ decay}

The theory of $0\nu\beta\beta$ decay was first formulated by Furry \cite{furry} and further developed by Primakoff and Rosen \cite{primakoff}, Molina and Pascual \cite{molina}, Doi \textit{et al.} \cite{doi81}, and, Haxton and Stephenson \cite{haxton}. Here we follow mainly the formulation of Tomoda \cite{tom91}. The phase space factors for $0\nu\beta\beta$ decay are simpler than those of $2\nu\beta\beta$ because of the absence of integration over the neutrino energies. Also, with two leptons in the final state and S-wave decay we can only form angular momentum $0$, $1$ and therefore the decay to $2^+$ is forbidden.

\subsubsection{$0^+\rightarrow 0^+_1$ $0\nu\beta\beta$-decay}
The differential rate for the decay is given by \cite{doi81,tom91}
\begin{equation}
\label{dw0nu}
dW_{0\nu}=\left(a^{(0)}+a^{(1)}\cos \theta_{12}\right)w_{0\nu}d\epsilon_1d(\cos \theta_{12})
\end{equation}
where $\epsilon_1$ and $\epsilon_2$ are the electron energies, $\theta_{12}$ the angle between the two emitted electrons, and 
\begin{equation}
w_{0\nu}=\frac{g_A^4 (G\cos\theta_C)^4}{16\pi^5}(m_e c^2)^2(\hbar c^2)(p_1c)(p_2c)\epsilon_1\epsilon_2
\end{equation}
This decay is forbidden by the standard model and can occur only if the neutrino has mass and/or there are right-handed currents. In view of recent experiments on neutrino oscillations \cite{SUP01, SNO2002, KAM2003} it appears that neutrinos have a mass and we therefore consider the phase space factors for this case. The quantities $a^{(0)}$ and $a^{(1)}$ in Eq.~(\ref{dw0nu}) can then be written as \cite{tom91}
\begin{equation}
a^{(i)}=f_{11}^{(i)}\left| \frac{\langle m_\nu\rangle}{m_e}\right|^2 \left| M_{0\nu}\right|^2
\end{equation}
$i=0,1$, where $M_{0\nu}$ is the nuclear matrix element and $f_{11}^{(0)}$, $f_{11}^{(1)}$ are the quantities given in Eq.~(\ref{combwave}). 

All quantities of interest are then given by integration of Eq.~(\ref{dw0nu}). Introducing 
\begin{equation}
F^{(i)}_{0\nu}=\frac{2}{\ln2}\int^{Q_{\beta\beta}+m_ec^2}_{m_ec^2}f^{(i)}_{11}w_{0\nu}d\epsilon_1,
\end{equation}
where $\epsilon_2$ is determined as $\epsilon_2=Q_{\beta\beta}+m_ec^2-\epsilon_1$, and defining the quantities
\begin{equation}
\label{g0nu}
G_{0\nu}^{(i)}=\frac{F^{(i)}_{0\nu}}{g_A^4(4R^2)}
\end{equation}
where $R=r_0A^{1/3}$, $r_0=1.2$ fm, is the nuclear radius, we can calculate:\\
(i) The half-life
\begin{equation}
\left[ \tau^{0\nu}_{1/2} \right]^{-1}=G_{0\nu}^{(0)}g_A^4 \left|\frac{\left\langle m_{\nu}\right\rangle}{m_e}\right|^2 \left| M_{0\nu}\right|^2,
\end{equation}
(ii) the single electron spectrum
\begin{equation}
\frac{dW_{0\nu}}{d\epsilon_1}={\cal N}_{0\nu} \frac{dG_{0\nu}^{(0)}}{d\epsilon_1}={\cal N}_{0\nu}\left[2f^{(0)}_{11}(\epsilon_1)w_{0\nu}(\epsilon_1)\right]
\end{equation}
where ${\cal N}_{0\nu}=g_A^4\left|\langle m_{\nu}\rangle/m_e\right|^{2}\left| M_{0\nu}\right|^2$.\\
(iii) and the angular correlation between the two electrons
\begin{equation}
\alpha(\epsilon_1)=\frac{f^{(1)}_{11}(\epsilon_1)}{f^{(0)}_{11}(\epsilon_1)}=\frac{dG_{0\nu}^{(1)}/d\epsilon_1}{dG_{0\nu}^{(0)}/d\epsilon_1}.
\end{equation}
The factor $(4R^2)$ has been introduced in Eq.~(\ref{g0nu}) to conform with standard notation \cite{boe92}, in which the nuclear matrix elements $M_{0\nu}$ are given in dimensionless units, that is they are multiplied by $R$. The factor of 4, which is missing in Tomoda's definition but is necessary to make the calculation consistent with Boehm and Vogel, has been the cause of considerable confusion in the literature, as well as the value of $r_0$ used in $R=r_0A^{1/3}$. Some authors use $r_0=1.1$~fm instead of $r_0=1.2$~fm.

\begin{figure}[h]
\includegraphics[width=0.65\linewidth ]{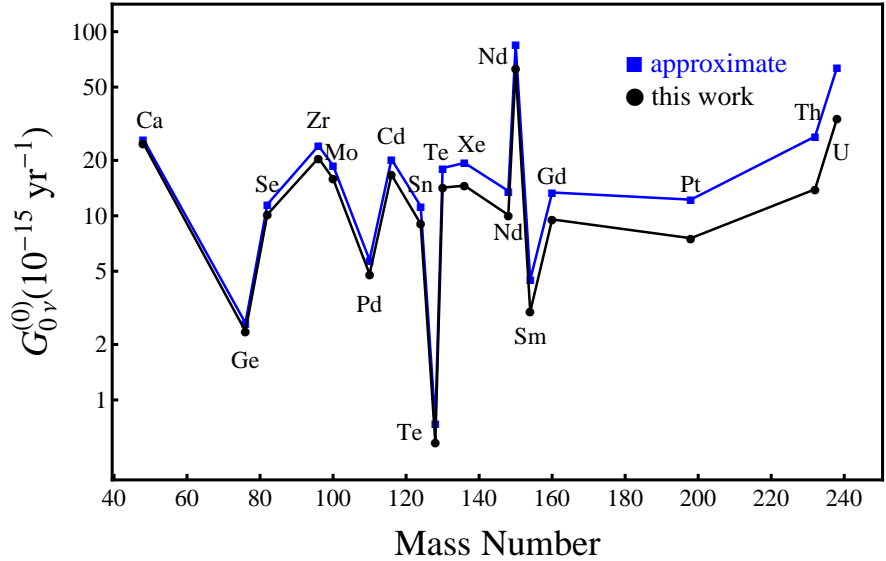} 
\caption{\label{0nuGcomp}Phase space factors $G_{0\nu}^{(0)}$ in units $(10^{-15}$ y$^{-1})$. The label "approximate" refers to the results obtained by the use of approximate electron wave functions.  The figure is in semilogarithmic scale.}
\end{figure}

We have done a calculation of $G_{0\nu}^{(0)}$ and $G_{0\nu}^{(1)}$ in the list of nuclei shown in Table~\ref{0nuG}. The obtained $G^{(0)}_{0\nu}$ values are also presented in Fig.~\ref{0nuGcomp} where they are compared with previous calculations \cite{boe92}.
\begin{ruledtabular}
\begin{center}
\begin{table}[h]
\begin{tabular}{lccc}
Nucleus 	&$G_{0\nu}^{(0)}(10^{-15}$ y$^{-1})$	&$G^{(1)}_{0\nu}(10^{-15}$ y$^{-1})$&$Q_{\beta\beta}$(MeV)
\cr \hline 
$^{48}$Ca	&24.81				&-23.09		&4.27226(404)\\
$^{76}$Ge	&2.363 				&-1.954		&2.03904(16)\\
$^{82}$Se	&10.16 				&-9.074		&2.99512(201)\\
$^{96}$Zr	&20.58 				&-18.67		&3.35037(289)\\
$^{100}$Mo	&15.92		 		&-14.25		&3.03440(17)\\
$^{110}$Pd	&4.815 				&-4.017		&2.01785(64)\\
$^{116}$Cd	&16.70				&-14.83		&2.81350(13)\\
$^{124}$Sn	&9.040				&-7.765		&2.28697(153)\\
$^{128}$Te	&0.5878 			&-0.3910	&0.86587(131)\\
$^{130}$Te	&14.22 				&-12.45		&2.52697(23)\\
$^{136}$Xe	&14.58 				&-12.73		&2.45783(37)\\
$^{148}$Nd	&10.10				&-8.506		&1.92875(192)\\
$^{150}$Nd	&63.03				&-57.76		&3.37138(20)\\
$^{154}$Sm	&3.015				&-2.295		&1.21503(125)\\
$^{160}$Gd	&9.559				&-7.932		&1.72969(126)\\
$^{198}$Pt	&7.556				&-5.868		&1.04717(311)\\
$^{232}$Th	&13.93				&-10.95		&0.84215(246)\\
$^{238}$U	&33.61				&-28.13		&1.14498(125)\\
\end{tabular}
\caption{\label{0nuG}Phase space factors $G_{0\nu}^{(0)}$ and $G^{(1)}_{0\nu}$ obtained using screened exact finite size Coulomb wave functions.}
\end{table}
\end{center}
\end{ruledtabular}

\begin{figure}[h]
\includegraphics[width=0.666\linewidth ]{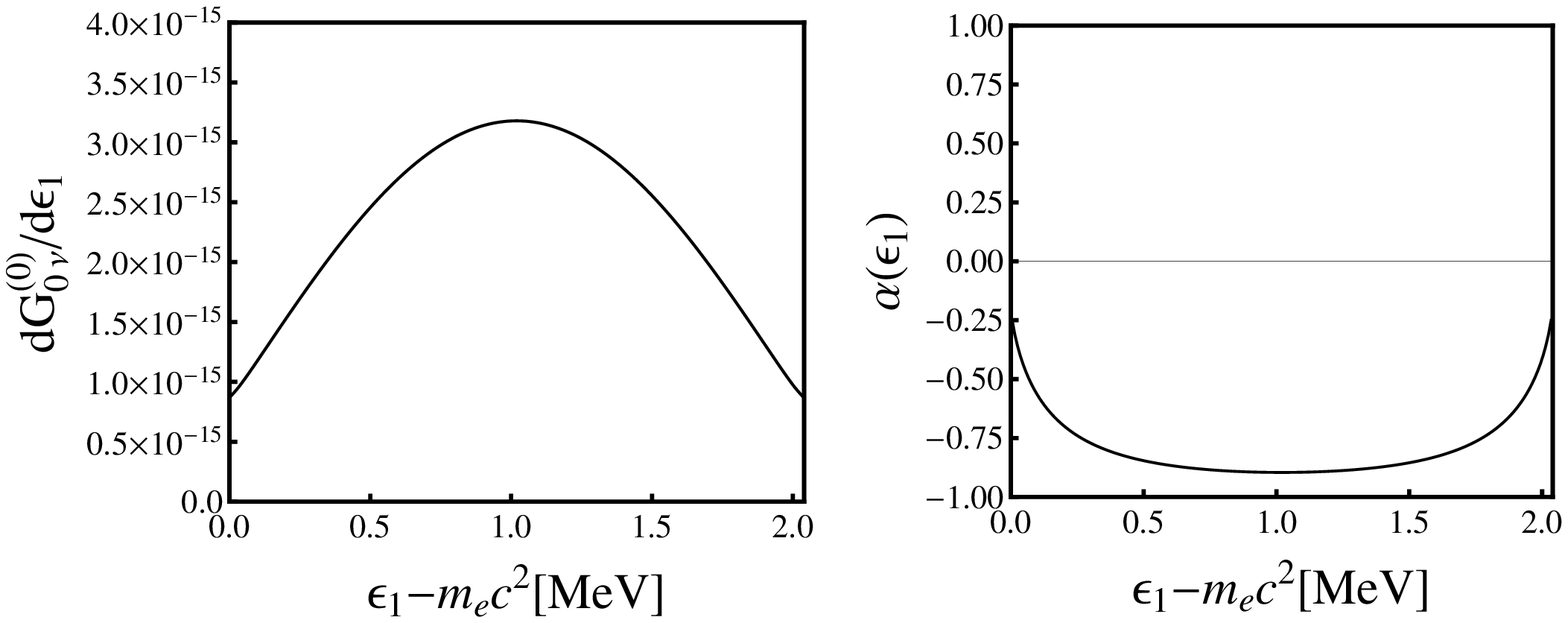} 
\caption{\label{Ge}Single electron spectra (left panel), and angular correlations between the two outgoing electrons (right panel) for the  $^{76}$Ge $\rightarrow ^{76}$Se $0\nu\beta\beta$-decay. The scale of the left panel should be multiplied by ${\cal N}_{0\nu}$ for a realistic estimate.}
\end{figure}

We also have available upon request the single electron spectra and angular correlation for all nuclei in Table~\ref{0nuG}. An example, $^{76}$Ge decay, is shown in Fig.~\ref{Ge}.

\subsubsection{$0^+\rightarrow 0^+_2$ $0\nu\beta\beta$-decay}
The decay to $0^+_2$ can also be calculated as in the previous subsection \ref{2nu02}. The results are shown in Table~\ref{0nuGex}.
\begin{ruledtabular}
\begin{center}
\begin{table}[h]
\begin{tabular}{lcccc}
Nucleus 	&$G_{0\nu}^{(0)}(10^{-15}$  y$^{-1})$	&$G^{(1)}_{0\nu}(10^{-15}$  y$^{-1})$ &$E(0^+_2)$(MeV) &$Q_{\beta\beta}(0^+_2)$(MeV)
\cr \hline 
$^{48}$Ca	&0.2989	 	&-0.2080	&2.99722(16) 	&1.27504(253)\\
$^{76}$Ge	&0.1776 	&-0.09855	&1.122283(7) 	&0.916757(167)\\
$^{96}$Zr	&4.566		&-3.760		&1.14813(7) 	&2.20224(296)\\
$^{100}$Mo	&3.162	 	&-2.493		&1.13032(10) 	&1.90408(27)\\
$^{110}$Pd	&0.08844	&-0.02958	&1.47312(12) 	&0.54773(76)\\
$^{116}$Cd	&0.7163		&-0.4075	&1.756864(24) 	&1.056636(154)\\
$^{124}$Sn	&0.1709		&-0.06237	&1.657283(22) 	&0.629687(1552)\\
$^{130}$Te	&0.3086	 	&-0.1271	&1.79352(11) 	&0.73345(34)\\
$^{136}$Xe	&0.6127	 	&-0.2924	&1.578990(23) 	&0.878840(393)\\
$^{148}$Nd	&0.2010		&-0.05354	&1.42446(4) 	&0.50429(196)\\
$^{150}$Nd	&27.27		&-23.26		&0.740382(22) 	&2.630998(222)\\
$^{154}$Sm	&0.2806		&-0.07744	&0.6806673(18) 	&0.5343627(12518)\\
$^{160}$Gd	&0.2063		&-0.04650	&1.279941(23) 	&0.449749(1283)\\
$^{232}$Th	&0.2622		&-0.0.1065	&0.69142(9)		&0.15073(255)\\
$^{238}$U	&0.7534 	&-0.03918	&0.94146(8)		&0.20352(133)\\
\end{tabular}
\caption{
\label{0nuGex}Same as Table~\ref{0nuG} but for the decay to the first excited $0^+$ state, $0^+_2$.}
\end{table}
\end{center}
\end{ruledtabular}

\section{Evaluation of the error}
\label{error}
The input parameters in the calculation of the phase space factors (PSF) are the $Q$-value, $Q_{\beta\beta}$, and the nuclear radius, $R$. We take the Q value from experiments whenever possible and thus the error introduced in $G$ is directly related to the experimental error. For example, recently the $Q$-value for $^{110}$Pd decay has been measured with high accuracy \cite{fin11}. Table~\ref{qerr} shows the improvement in the error in $G_{0\nu}^{(0)}$ and $G_{2\nu}^{(0)}$ due to the better accuracy obtained by measurement compared to the $Q$-value determined from mass values.
\begin{ruledtabular}
\begin{center}
\begin{table}[h]
\begin{tabular}{lcc}
$Q_{\beta\beta}$ keV 	&${G_{2\nu}}_{\rm{SSD}}^{(0)}$(y$^{-1}$)	&$G^{(0)}_{0\nu}$(y$^{-1}$)
\cr \hline 
2004.00(1133)$^{a}$			&$1.386(67)\times 10^{-19}$ 	&$4.707(86)\times 10^{-15}$ \\
2017.85(64)$^{b}$				&$1.469(05)\times 10^{-19}$  	&$4.815(06)\times 10^{-15}$ \\
\end{tabular}
\caption{\label{qerr}The uncertainty on PSF due to the uncertainty of the Q value. $^{a}$ From Ref.~\cite{ame03} and $^{b}$ from Ref.~\cite{fin11}.}
\end{table}
\end{center}
\end{ruledtabular}

The nuclear radius enters in the calculation in various ways, the most important of which is the evaluation of the quantities $g_{-1}(\epsilon)$ and $f_{1}(\epsilon)$. We evaluate the error here by comparing approximation (I) with (III) in a specific case, $^{110}$Pd, where the transition is $1g_{9/2}-1g_{7/2}$, obtaining an estimate of the error of $3\%$. For $0\nu$ decay the radius $R$ enters  also in the definition of $G_{0\nu}$. This is, however, an input parameter which does not depend on the method of calculation. We have used $R=r_0A^{1/3}$ with $r_0=1.2$ fm. We can estimate the error introduced by this choice by the same method used in the phase space factors for single-$\beta$ decay \cite{single}, that is by adjusting $r_0$ for each nucleus, $A$, $Z$, using 
\begin{equation}
\frac{3}{5}r_0^2A^{2/3}=\langle r^2\rangle _{exp},
\end{equation}
where $\langle r^2\rangle _{exp}$ is obtained from electron scattering and/or muonic x-rays. The largest difference between $R_{th}$ and $R_{exp}$ is found to be $\sim4\%$. This leads to an error estimate of 0.5\% for $2\nu$. For $0\nu$ we obtain an estimate of error of 7\%. 

In addition, we have an error coming from screening and most importantly from the value of $\langle E_N \rangle$. We estimate the screening error to be 10\% of the Thomas-Fermi contribution, known to overestimate the electron density at the nucleus. This gives an error in  $G_{0\nu}^{(0)}$, $G_{2\nu}^{(0)}$ of 0.1\%. The estimate of the error introduced by the choice of $\langle E_N \rangle$ is model dependent. If we vary $\tilde{A}$ from the value $1.12A^{1/2}$ MeV to the SSD value ($\sim 2$ MeV) we obtain for $^{110}$Pd decay an error of $7\%$, as shown in Fig.~\ref{atilde}b. If, however, we stay within a specific model, closure or SSD, the error estimate is much smaller. In particular for the SSD model the error is only arising from the value of $Q_{EC}$ and $Q_{\beta\beta}$ shown in Fig.~\ref{atilde}a. The estimate therefore depends on the nucleus considered. For $^{110}$Pd, the SSD model appears to be a good approximation and using it we obtain an estimate of the error of $0.05\%$. For the closure approximation the dependence of $G^{(i)}_{2\nu}$ on $\tilde{A}$ is very mild ($<1\%$) except very close to the threshold, $\left< E_N \right>=0$, as shown in Fig.~\ref{atilde}b. The situation is summarized in Table~\ref{unc}.
\begin{ruledtabular}
\begin{center}
\begin{table}[h]
\begin{tabular}{lcc}
$2\nu$ 	&$Q$-value	&$10 \times \delta Q/Q $\\
			&Radius 	&$0.5\%$ \\
			&Screening 	&$0.10\%$\\
			&$\langle E_N \rangle$ &model dependent\\
\cr \hline 
$0\nu$ 	&$Q$-value	&$3 \times \delta Q/Q $\\
			&Radius 	&$7\%$ \\
			&Screening 	&$0.10\%$\\
			&$\langle E_N \rangle$ &-\\
\end{tabular}
\caption{\label{unc}The estimate of uncertainties introduced to phase space factors $G_{2\nu}^{(0)}$ and $G_{0\nu}^{(0)}$ due to different input parameters.}
\end{table}
\end{center}
\end{ruledtabular}

\section{Use of phase space factor}
\begin{figure}[h]
\includegraphics[width=0.66\linewidth ]{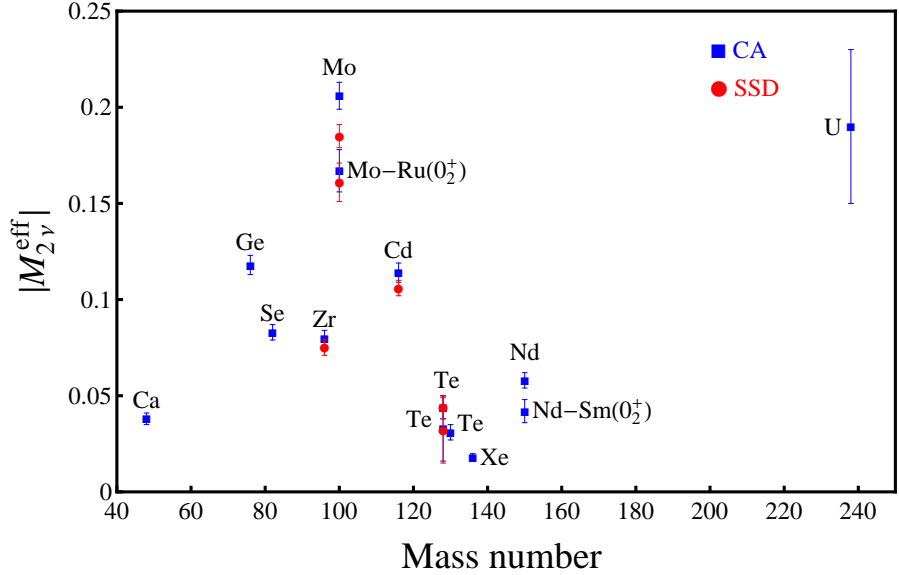} 
\caption{\label{Mefffig}Effective nuclear matrix elements $|M^{\rm{eff}}_{2\nu}|$ extracted from the experimental $2\nu\beta\beta$ half-lives as a function of mass number.}
\end{figure}

The main use of phase space factors (PSF) is in connection with a calculation of the nuclear matrix elements to predict life-times for the decay. Here an important point is that the nuclear matrix elements are defined in a way consistent with the phase space factors. For example, we have defined the phase space factors for $0\nu\beta\beta$ with a factor of 4 in Eq.~(\ref{g0nu}). This factor is not included in Tomoda's definition \cite{tom91} but it is in the book of Boehm and Vogel \cite{boe92}. The nuclear matrix elements consistent with this factor are, for GT, those of $\sum_{n,n'}\tau_n\tau_{n'}\vec{\sigma}_n\cdot\vec{\sigma}_{n'}$, not those of $(1/2) \sum_{n,n'}\tau_n\tau_{n'}\vec{\sigma}_n\cdot\vec{\sigma}_{n'}$.
We will present results of our predictions where phase space factors are combined with the IBM-2 nuclear matrix elements in a forthcoming publication \cite{bar11}. Here we use the calculation of PSF to extract the $2\nu$ matrix elements from experiments where the life-time of $2\nu\beta\beta$ decay has been measured. The quantity we extract is the dimensionless quantity $g_A^4|(m_ec^2)M_{2\nu}|^2=|M_{2\nu}^{\rm{eff}}|^2$ (also called ${\cal N}_{2\nu}$ in Sect.\ref{00 2bb-decay}). The extraction of $|M_{2\nu}^{\rm{eff}}|$ is possible in two cases: (1) the closure approximation (CA) and (2) the single state dominance (SSD) hypothesis. If neither of these two approximations is valid, then the quantities $G_{2\nu}$ and $M_{2\nu}$ cannot be separated as discussed after Eq.~(\ref{full}). The results obtained with the assumption of CA with $\tilde{A}=1.12A^{1/2}$ MeV and under the assumption of SSD for $^{96}$Zr, $^{100}$Mo, $^{116}$Cd, and $^{128}$Te are shown in Table~\ref{Meff} and in Fig.~\ref{Mefffig}.
We note that all effective matrix elements in Table~\ref{Meff} vary between a minimum of $\sim 0.02$ ($^{136}$Xe) and a maximum of $\sim 0.2$ ($^{100}$Mo and $^{238}$U), with the majority being $\sim 0.05$.
\begin{ruledtabular}
\begin{center}
\begin{table}[h]
\begin{tabular}{lccccc}
Nucleus   &$G_{2\nu}^{(0)}(10^{-21}$y$^{-1})$ &${G_{2\nu}^{(0)}}_{\rm{SSD}}(10^{-21}$ y$^{-1})$  	&$\tau_{1/2}^{2\nu}(10^{18}$ y) exp$^a$ &$|M^{\rm{eff}}_{2\nu}|$ &$|M^{\rm{eff}}_{2\nu}|_{\rm{SSD}}$\cr \hline
$^{48}$Ca	&15550. & 	&$44^{+6}_{-5}$		&$0.038\pm0.003$	&	\\
$^{76}$Ge	&48.17 	&	&$1500\pm100$		&$0.118\pm0.005$ 	&	\\
$^{82}$Se	&1596. 	& 	&$92\pm7$			&$0.083\pm0.004$ 	&	\\
$^{96}$Zr	&6816. 	&7825. 	&$23\pm2$			&$0.080\pm0.004$ 	&$0.075\pm0.004$	\\
$^{100}$Mo	&3308. 	&4134. 	&$7.1\pm0.4$		&$0.206\pm0.007$ 	&$0.185\pm0.006$	\\
$^{100}$Mo-$^{100}$Ru($0^+_2$)	&60.55 &65.18 	&$590^{+80}_{-60}$	&$0.167\pm0.011$	&$0.161\pm0.010$\\
$^{116}$Cd	&2764. 	&3176.	&$28\pm2$			&$0.114\pm0.005$ 	&$0.106\pm0.004$	\\
$^{128}$Te	&0.2688 &0.2727	&$1900000\pm400000$	&$0.044\pm0.006$ 	&$0.044\pm0.006$	\\
$^{128}$Te	&0.2688 &0.2727 &$3500000\pm2000000^b$&$0.033\pm0.017$ 	&$0.032\pm0.017$	\\
$^{130}$Te	&1529.  &	&$680^{+120}_{-110}$&$0.031\pm0.004$ 	&	\\
$^{136}$Xe	&1433.  &	&$2110\pm 250^c$	&$0.0182\pm0.0017$	&	\\
$^{150}$Nd	&36430. & 	&$8.2\pm0.9$		&$0.058\pm0.004$	&	\\
$^{150}$Nd-$^{150}$Sm($0^+_2$)	&4329. & 	&$133^{+45}_{-26}$	&$0.042\pm0.006$	&	\\
$^{238}$U	&14.57 	& 	&$2000\pm 600$		&$0.19\pm0.04$		&		\\
\end{tabular}
\caption{\label{Meff}Experimental $2\nu\beta\beta$ half-lives and the corresponding effective nuclear matrix elements $|M^{\rm{eff}}_{2\nu}|$. For the case $^{128}$Te two experimental half-lives are listed, upper from evaluation of Barabash $^a$\cite{bar10} and the lower from the comment of Pritychenko $^b$\cite{pri10}. The value for $^{136}$Xe is from a new measurement and is taken from $^c$\cite{exo}. 
}
\end{table}
\end{center}
\end{ruledtabular}

The effective matrix elements $|M^{\rm{eff}}_{2\nu}|^{\rm{exp}}$ can, in principle, be obtained from measurements of $GT^{\pm}$ strengths (and $F^{\pm}$ strengths), through the formula
\begin{equation}
\label{mgt}
M^{GT}_{2\nu} =\sum_N \frac{\langle 0^+_F||\tau^+\vec{\sigma} ||1^+_N\rangle \langle 1^+_N||\tau^+\vec{\sigma} ||0^+_I\rangle}{\frac{1}{2}(Q_{\beta\beta}+2m_ec^2)+E_N-E_I},
\end{equation} 
$\left| M^{\rm{eff}}_{2\nu} \right|^{\rm{exp}}=g_A^2 \left| (m_ec^2) M^{GT}_{2\nu} \right|$, and similar formulas for the Fermi matrix elements. However, in experiments, only the magnitude of the individual GT matrix elements can be measured, not its sign. Furthermore, it must be decided to what $N$ to stop the evaluation of the sum, and what value to use for $g_A$. Therefore, theoretical models must be used to obtain $\left| M^{\rm{eff}}_{2\nu} \right|$ for $GT^{\pm}$ strengths. A recent example is $^{150}$Nd decay \cite{gue11} (one should note that in this paper the denominator in the definition of $2\nu$ Gamow-Teller nuclear matrix element is different from Eq.~(\ref{mgt}) by $2m_ec^2$ due to the use of atomic masses in the calculation of $Q_{\beta\beta}$ and $E_N-E_I$), where $M^{GT}_{2\nu}$ has been extracted under (i) the assumption that only the $1^+$ state at $E_x$($^{150}$Pm)$=0.11$~MeV contributes to the decay and (ii) that all states up to $E_x$($^{150}$Pm)$<3.0$~MeV contribute. The result is (i) $M^{GT}_{2\nu}$(MeV$^{-1}$)$=0.028\pm 0.006$ and (ii) $M^{GT}_{2\nu}$(MeV$^{-1}$)$=0.13\pm 0.02$. Multiplying by $(m_ec^2)=0.511$ MeV and $g_A^2=1.273^2$ \cite{mar11}, one obtains (i) $\left| M^{\rm{eff}}_{2\nu}\right|^{\rm{exp}}_{\rm{SSD}}= 0.023\pm 0.005$ and (ii) $\left| M^{\rm{eff}}_{2\nu}\right|^{\rm{exp}}= 0.108\pm 0.017$. These two estimates bracket our extracted value $0.058\pm 0.004$. This "experimental" way of extraction also assumes that the factorization of $\tau^{2\nu}_{1/2}$ to  $G_{2\nu}$ and $M_{2\nu}$ is valid.

Our calculation of ${G_{2\nu}^{(0)}}_{\rm{SSD}}$ allows one to test the SSD assumption for $^{100}$Mo, $^{116}$Cd, and $^{128}$Te, where the matrix elements (even-even $\to$ odd-odd) $0^+ \to1^+_1$ and (odd-odd $\to$ even-even) $1^+_1 \to 0^+$ are known from single $\beta$ decay experiments. The extracted values of $\left| M^{\rm{eff}}_{2\nu} \right|^{\rm{exp}}$ using the single $\beta$ decay (or EC) matrix elements and $g_A=1.273$ are $\left| M^{\rm{eff}}_{2\nu} \right|^{\rm{exp}}=0.174\pm0.075$, $0.148\pm0.023$, $0.0152\pm0.0003$ for $^{100}$Mo, $^{116}$Cd, and $^{128}$Te, respectively. These values, as well as that of $^{150}$Nd discussed above, are given in Table~\ref{meffexp} and compared with the ones obtained from experimental double $\beta$ decay half-lives. The SSD model appears to give a rather good agreement for $^{100}$Mo and $^{116}$Cd, but is off by a factor of 2 in defect in $^{128}$Te and $^{150}$Nd. The situation has been also analyzed in detail from different point of view in Ref.~\cite{eji10}.

\begin{ruledtabular}
\begin{center}
\begin{table}[h]
\begin{tabular}{lccc}
Nucleus   &$|M^{\rm{eff}}_{2\nu}|$ &$|M^{\rm{eff}}_{2\nu}|_{\rm{SSD}}$ &$|M^{\rm{eff}}_{2\nu}|^{\rm{exp}}_{\rm{SSD}}$\cr \hline
$^{100}$Mo						&$0.206\pm0.007$ 	&$0.185\pm0.006$	&$0.174\pm0.075$\\
$^{100}$Mo-$^{100}$Ru($0^+_2$)	&$0.167\pm0.011$	&$0.161\pm0.010$ 	&$0.104\pm0.045$\\
$^{116}$Cd						&$0.114\pm0.005$ 	&$0.106\pm0.004$	&$0.148\pm0.023$\\
$^{128}$Te						&$0.044\pm0.006$ 	&$0.044\pm0.006$	&$0.0152\pm0.0003$\\
$^{128}$Te						&$0.033\pm0.017$ 	&$0.032\pm0.017$	&$0.0152\pm0.0003$\\
$^{150}$Nd						&$0.058\pm0.004$	&					&$0.023\pm0.005$\\
\end{tabular}
\caption{\label{meffexp}Effective nuclear matrix elements $|M^{\rm{eff}}_{2\nu}|$, $|M^{\rm{eff}}_{2\nu}|_{\rm{SSD}}$ obtained from experimental $2\nu\beta\beta$ half-lives compared with $|M^{\rm{eff}}_{2\nu}|^{\rm{exp}}_{\rm{SSD}}$, the effective nuclear matrix elements obtained from single $\beta$ decay experiments ($^{100}$Mo, $^{116}$Cd, $^{128}$Te), or from $GT^{\pm}$ strength measurements ($^{150}$Nd). For the case $^{128}$Te the two values listed are explained in the caption of Table~\ref{Meff}. 
}
\end{table}
\end{center}
\end{ruledtabular}

\section{Conclusions}
In this article, we have reported a complete and improved calculation of phase space factors for $2\nu\beta^-\beta^-$ and $0\nu\beta^-\beta^-$ decay, including half-lives, single electron spectra, summed electron spectra, and electron angular correlations, to be used in connection with the calculation of nuclear matrix elements. Apart from their completeness and consistency of notation, we have improved the calculation by using exact Dirac wave function with finite nuclear size and electron screening. The program for calculation of phase space factors has been set up in such a way that additional improvements may be included if needed (P-wave contribution, finite extent of nuclear surface, etc.) and that it can be used in connection with the closure approximation, the single state dominance hypothesis and the calculation with sum over individual states. 
In a subsequent publication we are planning to present complete and improved calculations for $2\nu\beta^+\beta^+$ and $0\nu\beta^+\beta^+$ decay, as well as of the competing processes $2\nu\beta^+EC$, $2\nu ECEC$ and $0\nu\beta^+EC$, $0\nu ECEC$.

\begin{acknowledgments}
This work was performed in part under the US DOE Grant DE-FG-02-91ER-40608. We wish to thank all the experimental groups that have stimulated our work in particular A. Bettini, S. Elliott, E. Fiorini, G. Gratta, A. McDonald, S. Schoenert and K. Zuber.
\end{acknowledgments}

\end{document}